# Modeling the Effect of Data Redundancy on Speedup in MLFMA Near-Field Computation

Morteza Sadeghi

University of Tehran, sadeghi.morteza@ut.ac.ir

## Abstract

The near-field (P2P) operator in the Multilevel Fast Multipole Algorithm (MLFMA) is a performance bottleneck on GPUs due to poor memory locality. This work introduces data redundancy to improve spatial locality by reducing memory access dispersion. For validation of results, we propose an analytical model based on a *Locality* metric that combines data volume and access dispersion to predict speedup trends without hardware-specific profiling. The approach is validated on two MLFMA-based applications: an electromagnetic solver (DBIM-MLFMA) with regular structure, and a stellar dynamics code (PhotoNs-2.0) with irregular particle distribution. Results show up to 7× kernel speedup due to improved cache behavior. However, increased data volume raises overheads in data restructuring, limiting end-to-end application speedup to 1.04×. While the model cannot precisely predict absolute speedups, it reliably captures performance trends across different problem sizes and densities. The technique is injectable into existing implementations with minimal code changes. This work demonstrates that data redundancy can enhance GPU performance for P2P operator, provided locality gains outweigh data movement costs.

## 1. Introduction

The Multilevel Fast Multipole Algorithm (MLFMA) [2][3] is a cornerstone for accelerating large-scale N-body problems in computational electromagnetics [50], astrophysics [46], and other potential-based simulations.

While the far-field operator of MLFMA has been extensively optimized for CPU clusters through hybrid, hierarchical and ternary parallelization [19][14][15][16] and heterogeneous clusters [17][18], the near-field (P2P) operator -computationally similar to the original N-body problem- remains a performance bottleneck on single GPUs. Given that GPU implementations offer superior energy efficiency and performance for N-body workloads [8], especially when leveraging low-level optimizations [5] and memory-centric techniques like thread tiling and data layout tuning [6][7], we focus on accelerating the P2P operator using CUDA. Our data redundancy approach enhances thread-level independence, making it compatible with existing hybrid or distributed MLFMA frameworks where reduced memory contention benefits large-scale GPU cluster execution.

Near-Field matters more in N-body problems -e.g., chemical simulations [57], stellar dynamics [45], electrostatics [53] -where mesh Ewald or single-level FMM is used. These involve sparse particle distributions in boxes during long MLFMA simulations, needing parallelization distinct from EM kernels (which handle dense, regular 3D meshes).

Yokota et al. tackled time-evolving particle systems on GPUs using FMM variants, focusing on throughput and reduced data movement/sync via task grouping for unbalanced FMM trees in molecular dynamics. Their go-to methods: level-synchronous tree traversal and coarse-grain task parallelism with manual GPU memory control. In [56], they sped up a tree-code FMM [55] on a single GPU by batching kernels -queuing interactions, reorganizing data into SoA arrays, then executing the whole queue -since many boxes are empty or workloads uneven, improving speed and memory coalescing. In [57] they scaled this to GPU clusters using kernel batching, zero-padding for better coalescing, and node-level interaction distribution with redundant data (like [54]) to cut inter-node dependencies and communication. They also cut redundant symmetrical interaction computations, lowering GPU task wait times. In [58] they studied FMM's P2P operator on many-core systems, testing task granularity (outer vs. inner loop parallelization); inner loops gave better cache efficiency for small problems, revealing underused cache potential in P2P. One notable finding in this work is that MPI communication in the P2P phase -specifically for neighbor access -is best improved with a topology-aware approach: MPI tasks exchange data only with relevant (nearby) tasks, not with all tasks. Dehnen's [59] FMM for stellar dynamics was recently combined with Yokota's [55] parallelization in [60], running on heterogeneous setups (two GPU types) via OpenCL. They tested memory-write conflict strategies, showing result duplication + reduction beats lock-based methods, yielding higher speedup and more compute divergence. [46], another FMM-based stellar dynamics sim, uses a similar approach. [60] also assessed batch size and loop unrolling for resource use. In [62],[63], Yokota's group tested Laplace kernel [63] scalability for P2P, concluding P2P isn't memory-bound -a claim echoed in [64]. They also found MLFMA load-balances well with space-filling curves (localized data access, no far jumps). Yet tree building and data communication remain FMM parallelization bottlenecks for N-body apps. Despite this, auto-parallelization frameworks like GENGA [7] and SYCL studies [8] confirm irregular neighbor access in P2P is still a main GPU bottleneck -even with reduced algorithmic complexity.

We argue that the P2P operator in MLFMA is equally critical and under-optimized -or, more cautiously, holds significant potential for improved GPU performance. This work shifts focus to optimizing the P2P operator on a single GPU, where performance is often degraded by memory access patterns. The P2P phase involves direct particle-to-particle interactions within neighboring cells, forming a stencil-like computation where each processing unit accesses data from multiple adjacent and non-adjacent memory blocks. This leads to poor spatial data locality, high cache miss rates, and inefficient global memory bandwidth utilization-critical issues on GPU architectures which are sensitive to memory hierarchy behavior [35][26] ans is common in P2P operator in sparse particle distribution problems [58][7][8].

## 1.1. The Missing FrameWork

Prior GPU optimizations for stencil and N-body codes emphasize techniques such as thread tiling, sharing-memory usage [6], and data layout transformation (e.g., Structure-of-Arrays (SoA) to improve coalescing) [7]. For MLFMA's P2P, some works have explore on-the-fly data computation and load data to registers to reduce memory footprint [9] decoupling self-box and neighbor-box interactions to improve memory coalescing [10] and hierarchical neighbor lists for efficient search in large scale imbalanced trees [17] which are of known techniques for stencil optimization [25][26].

While many MLFMA optimizations are developed intuitively and validated after implementation, there is no method to predict the impact of algorithmic changes—such as data restructuring—on performance before coding. Most approaches rely on trial and error, without separating the effects of individual optimization steps. A prior work by [19][15] presents a conceptual parallelization framework for MLFMA in electromagnetic simulations, focusing on task distribution for CPU clusters. While useful, it stays at an abstract level and does not quantify performance gains from specific algorithmic changes. Moreover, earlier works have not explored data-driven techniques like data layout modifications or memory access optimizations, which can be critical for speedup.

This creates an opportunity to introduce such methods. We focus on data-driven solutions as a form of algorithm-agnostic optimization—approaches that do not modify the core computation but instead restructure data to improve performance on shared-memory devices like GPUs. We specifically study data redundancy as a representative restructuring technique and propose a simple analytical model to estimate its impact on end-to-end speedup. Our approach models how restructuring affects each phase of execution: data collection, data transfer to GPU, kernel computation, and result update.

The model is built on propositions derived from prior GPU performance studies [35-42] and provides a predictive formula for speedup. It operates at a high level of abstraction, is vendor-independent, and relies only on algorithmic features present in the data structure—such as clustering threshold or mesh density in last level boxes in P2P operator. This model separates the modeling from GPU architecture, algorithm design, parallelization strategy, and implementation details. As a result, our model is interpretable, does not require profiling, and can be applied statically during early development. Furthermore, by focusing on a single operator—the P2P component of MLFMA—our approach is kernel-independent and requires minimal code changes for integration. Without such a model, it is difficult to determine whether data restructuring will improve or degrade overall performance. However, we acknowledge a trade-off: while the model is practical and generalizable, it is less accurate in predicting absolute speedups and works best for capturing performance trends in clean, well-structured MLFMA implementations.

### 1.1. Data Restructuring via Redundancy: A Promising but Underexplored Strategy

Prior GPU optimizations for stencil and N-body codes emphasize techniques such as thread tiling, sharing-memory usage [6], and data layout transformation (e.g., Structure-of-Arrays (SoA) to improve coalescing) [7]. For MLFMA's P2P, some works have explore on-the-fly data computation and load data to registers to reduce memory footprint [9] decoupling self-box and neighbor-box interactions to improve memory coalescing [10] and hierarchical neighbor lists for efficient search in large scale imbalanced trees [17] which are of known techniques for stencil optimization [25][26].

However, prior P2P approaches [10][11][12][17] do not fundamentally address the spatial dispersion of memory accesses. In contrast, data redundancy -duplicating neighbor data per thread or block -offers a direct path to enhancing spatial locality. By replicating data in global memory, threads can access all required neighbor information from a compact, contiguous region, reducing the number of non-adjacent memory fetches and cache misses.

Although data redundancy introduces overhead in memory footprint and data transfer, we hypothesize that the reduction in cache misses on GPU can yield a net performance gain. This is especially plausible in sparse or irregular problems (e.g., stellar dynamics [44], electrostatics) where redundant data still fits within cache capacity. Data redundancy is also inherently used in Yokota's task-based approach, where cell data is duplicated per interaction in the task queue [57]. They identified tree building (i.e., data collection) as a major bottleneck -an overhead we also observed in our results. Our approach also addresses the P2P write-conflict problem using a solution similar to [60]: duplicating results and performing a reduction afterward through the update process.

### 1.2. Need for Predictive Modeling Before Implementation

Applying data redundancy involves trade-offs: increased data volume vs. improved locality. Blindly implementing it risks performance degradation due to memory bottlenecks or cache thrashing. Therefore, a predictive model is essential to evaluate the potential benefit before code modification.

While a few prior works have modeled MLFMA performance, such as [12][13], which model single-level FMM on heterogeneous systems using a key "Group Size" parameter, their models rely on hardware-dependent coefficients derived from profiling. Similarly, machine learning-based models [27][28] offer high accuracy but act as black boxes, lack interpret-ability, and require extensive training data-making them unsuitable for early-stage design decisions.

We argue that an analytical, interpret-able, and architecture-independent model is needed to guide the application of data restructuring. Such a model should capture the algorithmic impact of redundancy-on memory access patterns, data volume, and computational complexity-without requiring system-specific profiling.

### 1.3. Modeling Speedup via Data Locality

We propose a fully analytical framework to model the speedup of GPU-based MLFMA applications under data redundancy. Our approach centers on a custom metric called Locality, which quantifies the compactness of memory accesses by combining: Data Access Dispersion (D): number of non-adjacent memory blocks accessed by a thread, Volume (V): total data volume accessed per thread and Reusability (R): reuse of fetched data within a thread. We have no idea for measuring he last one and instead try to consider its effects in piece-wise formulation.

We define Locality as a function of how restructuring alters these factors. The core insight is that improved locality reduces cache misses, thereby accelerating kernel execution. To make the model practical and injectable into existing implementations, we limit restructuring to the P2P operator. The overall speedup is modeled as a composition of changes in data collection, GPU data transfer, kernel computation and update data based on results. Furthermore, our data redundancy approach enhances sub-problem independence by localizing memory accesses, making it naturally compatible with existing hybrid or cluster-based MLFMA frameworks -where improved thread-level autonomy reduces inter-node synchronization pressure and facilitates integration into large-scale GPU-accelerated systems.

## 2. Related Works

### 2.1. MLFMA on GPUs: Far-Field vs. Near-Field Optimization

The parallelization of MLFMA on GPU clusters has matured significantly [14], with efforts focusing on far-field acceleration through algorithmic modifications [15][16] and efficient distribution policies [17]. On a single GPU, far-field computation has also been optimized for better occupancy and memory use [21].

In contrast, the near-field (P2P) operator has seen limited GPU-specific optimization. Some works execute P2P naively on GPU or even offload it to CPU [9]. Others propose on-the-fly computation [9] to reduce memory transfer, decomposition of interactions [10] to improve coalescing and combined techniques [11] achieving high speedups in specific applications like ray tracing (NGIM). These methods improve performance but do not explicitly model or optimize for spatial data locality, which is central to our approach.

### 2.2. Stencil Optimization

The P2P operator exhibits a stencil-like access pattern (order-1-box stencil), and general GPU stencil optimizations [25][26] are relevant in prior works like thread tiling and sharing-memory usage [6][11], block merging [50] and SoA vs. AoS data layouts [7].

SoA improves coalescing for per-field access but increases the number of memory instructions per thread. AoS bundles fields, reducing instruction count but potentially increasing cache pressure. Our data redundancy technique can be viewed as an AoS-like restructuring with duplication, enhancing locality at the cost of increased volume.

### 2.3. Taske-based Approaches

Additionally, task-based parallelization approaches have been proposed to improve load balancing and scalability in MLFMA, particularly on heterogeneous architectures. Works such as [22][23][24] introduce dynamic runtime systems that decompose MLFMA operations into fine-grained tasks, enabling efficient scheduling across multi-core and GPU nodes. In [65] task grouping with level based synchronous execution is introduced for MLFMA as an efficient and scalable parallelization technique for solving large scale problems with imbalanced trees (in levels and particle distribution) on clusters. These methods enhance parallelism and resource utilization but do not explicitly address memory access locality in the near-field computation. They can improve efficiency of near-field operator based on task granularity and algorithmic parameters, however is not of our interest since their analysis and modeling requires a different modeling methodology.

## 2.4. Data Restructuring and Redundancy in Prior Work

Data restructuring for performance has been explored in prior works like matrix layout optimization for chain multiplication yielded 1.8X speedup [32], bandwidth-aware sampling to minimize GPU transfers in GNNs [33] and graph partitioning to reduce inter-node communication in AIM-MLFMA [34].

Compiler-level techniques have also proposed data replication in caches [40][43], but these are limited to simulation or specific runtime conditions.

Our work differs by applying explicit, algorithm-aware data redundancy in global memory to enhance locality in stencil-based MLFMA kernels-a strategy not previously modeled or widely applied.

## 2.5. GPU Performance Modeling: From Profiling to Prediction

Modeling GPU performance remains challenging due to micro-architectural complexity. Existing approaches include hybrid analytical + ML models [27][28] which combine static features with dynamic profiling but require extensive data, cycle-accurate simulators [29] which are precise but slow and architecture-specific, application-specific models for FMM [12][13] that rely on hardware coefficients.

Crucially, prior research in GPU performance modeling has consistently identified data locality and cache misses as dominant factors in kernel performance. Studies such as [35][36][37][41] have quantitatively demonstrated that poor spatial locality leads to high cache miss rates, which in turn impose severe penalties due to long memory latencies on GPUs. The spatial locality is the same important in many-core architectures [38]. [39] further identifies cache miss events as one of the top performance-limiting factors in GPU applications, especially for memory-bound kernels. [29] counts the number of memory accesses by tracking the code execution and uses it as a metric for data access latency rather than simulating behavior of MSHR registers and DRAM queue. Moreover, modeling efforts like [27][28] acknowledge that memory hierarchy effects-particularly cache behavior and data reuse-are central to accurate performance prediction. These works incorporate dynamic profiling or machine learning to capture non-linear memory effects, implicitly confirming that locality is not just intuitive but measurable and impactful.

Our model diverges by being purely analytical and device-independent, focusing on algorithmic parameters and data access patterns rather than hardware features. We avoid profiling by using time complexity and static locality analysis to predict speedup trends.

## 2.6. Modeling Data Transfer

In overall, the precise data transfer modeling for small volume kernels is a complicated task which is solved by profiling the application and using machine learning models [27][28] but for large volume kernels is analytical but dependent on architecture parameters and could be done by static analysis of code [27-29]. In [1] exceptionally, the data transfer was modeled as a linear coefficient for a vendor specific case of typical electrostatic MLFMA application. Since In our work the speedup is targeted instead of running time, we are able to simplify data transfer model and making it device independent by considering only the data volume.

## 3. The Proposed Modeling Framework

We model the speedup of restructured version compared with baseline (existing) implementation instead of execution time or instructions per cycle for that it is more simple to evaluate the ratio of major factors affecting execution time rather than accurately calculation of IPC and comparing them. In the case that the work doesn't have a GPU implementation, we propose a GPU version and compare restructuring with the proposed version. A schematic of the proposed framework is shown in Figure [1].

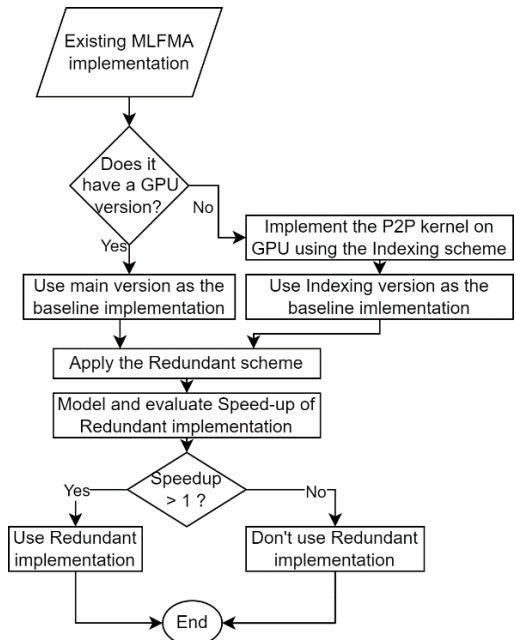

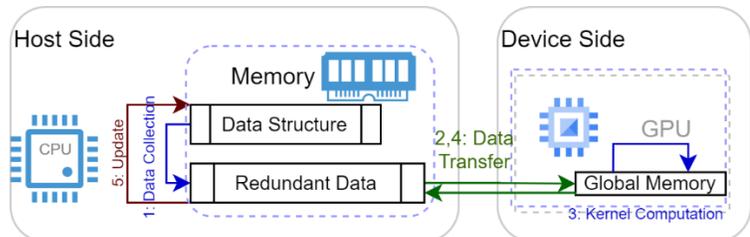

Figure[1] Flowchart for applying the proposed framework to existing MLFMA implementations.

Figure[2] Runtime components and Setp Environment.

### 3.1. Hypothesis and Core Idea

**Hypothesis:** Duplicating data in GPU global memory reduces cache misses by improving spatial locality, leading to net speedup despite increased data volume.

We model the effect of data restructuring -specifically redundancy -on the end-to-end speedup of MLFMA applications.

### 3.2. Key Propositions

**Proposition 1 (Amdahl's Law with Restructuring):** The total speedup is:

$$X_{total} = \frac{serial_{base} + parallel_{base}}{serial_{rest} + parallel_{rest}} \qquad (1)$$

where base refers to base algorithm and rest refers to the version with data restructuring, and serial and parallel refers to serial and parallel code sections. Data restructuring affects both serial and parallel sections. Our model aims to predict this speedup.

**Proposition 2 (Total Execution Time):** The total time of the modified algorithm includes four components:
$$T_{total} = T_{Collect} + T_{Transfer} + T_{Compute} + T_{Update} \quad (2)$$
$$T_{transfer} = T_{transfer}GPU + T_{transfer}GPU \quad (3)$$
A schematic of the runtime components is shown in Figure[2].

**Proposition 3 (Speedup of Modified Component):** The speedup of a restructured component without considering other components is:
$$X_{op} = \frac{1}{\left(\frac{share_{op}}{X_{op}}\right) + (1 - share_{op})} \quad (4)$$
where $X_{op}$ is the speedup of the modified operation, and $Share_{op}$ is its time share in the baseline application.

Combing formulas 1-4 we define the final speedup as :
$$X_{rest} = \frac{1}{\left(\frac{share_{p2p}}{X_{p2p}}\right) + (1 - share_{p2p})} \quad (5)$$
$$X_{p2p} = \frac{1}{\left(\frac{share_{collect}}{X_{collect}}\right) + \left(\frac{share_{transfer}}{X_{transfer}}\right) + \left(\frac{share_{compute}}{X_{compute}}\right) + \left(\frac{share_{update}}{X_{update}}\right)} \quad (6)$$

### 3.3. Modeling Individual Components

**Conjecture 1 (Memory Operations):** Data collection and update times scale with time complexity and data volume:
$$Speedup \propto X_{Complexity} \times X_{mem}, X_{Complexity} = \frac{Complexity_{base}}{Complexity_{restr}}, X_{mem} = \frac{V_{base}}{V_{restr}} \quad (7)$$

**Proposition 4 (Data Transfer):** Transfer time is linear in data volume:
$$X_{transfer} \propto \frac{V_{base}}{V_{restr}} \quad (8)$$

**Theory 1 (GPU Computation Speedup):** Kernel speedup depends on time complexity, kernel launches, and locality. We use term kernel launch here like [1] to refere the number of times that kernel must use all of cores to serve all CUDA threads.
$$X_{compute} = X_{ComplexityKernel} \times X_{Locality} \times X_{KernelLaunch} \quad (9)$$

### 3.4. Definition of Locality

We define *Locality* based on memory access behavior as follows:

**Proposed Definition 1:** For a thread, let D be the number of non-adjacent memory blocks accessed, and V the data volume. If data restructuring, here specifically the redundancy, reduced the data dispersion and the restructured data fits in cache (V≤C ), then we expect increase in reusability and reduce in cache misses hence increase in speedup :

$$V \leq C \rightarrow Cachemiss \propto 1/(D' \cdot V'), X_{Locality} \propto D' \cdot V' \tag{10}$$

where $D', V'$ are ratio of dispersion and volume of Indexing to Redundant respectively. If redundant data exceeds cache (V>C) then it reduces reusability therefore we must consider its negative impact:

$$V > C \rightarrow Cachemiss \propto D'/V', X_{Locality} \propto V'/D' \tag{11}$$

Thus, redundancy improves speedup when it reduces dispersion and the total volume fits in cache. We don't have any offer for quantifying the data reusability, instead we combined it with dispersion and cache capacity as an intuitive concept.

### 3.5. When is Redundancy Effective?

We mentioned above that data redundancy as a kind of restructuring might increase number of cache misses and reduce reusability chance of cache data. This guides us to introduce higher levels of redundancy where data is duplicated per coarser processing elements, means duplicating data per thread block rather than threads. Hence for Low particle density, sparse distribution we offer Thread-level redundancy while for High density, regular distribution we offer Block-level redundancy.

The efficiency of redundancy grain might be entangled with task distribution grain. For Uniform particle distribution *Thread-per-particle* tasking is more popular while for Irregular distribution, *Thread-per-box* tasking seems more reasonable. This limits the choice for redundancy level and make it less applicable.

For P2P operator it could be said that the redundancy is not effective when a small fraction of the runtime is spent on near-field computations, there are a lot of highly dense boxes and with high non-uniform data distribution or a lot of small boxes with high data volume, or when the redundanted data desn't fit in GPU memory or imposes more cache misses.

### 3.6. Advantages and Limitations

Our modeling technique is simple to implement in existing MLFMA codes, analytical and interpretable -provides insight into performance drivers, and architecture-independent and requires no profiling. Although it is not an automated tool for computing Locality and relies on manual analysis, its accuracy is lower than ML or simulation-based models, it focuses on global memory; assumes sharing-memory/register usage is consistent, it is evaluated only on single-GPU systems, and finally it does not predict absolute speedup precisely, but captures trends reliably.

## 4. System Setup and Algorithm Selection

To evaluate the impact of data redundancy on GPU performance in real-world MLFMA applications, we selected two representative works that differ fundamentally in problem structure, data distribution, and computational workflow. This selection challenges the strength of modeling framework and applicability of data redundancy.

**DBIM-MLFMA [50]**: A GPU-accelerated solver for inverse electromagnetic scattering using the Distorted Born Iterative Method (DBIM). This application operates on a uniform mesh of a solid, static object, leading to a regular quad-tree structure in MLFMA. The MLFMA is used repeatedly within an iterative optimization loop (gradient descent), where the matrix-vector product is computed. Crucially, the geometry and tree structure remain fixed across iterations, enabling pre-computation and reuse of interaction patterns. While originally designed for large GPU clusters [51], we adapted their published code [49] for a single-GPU setting to isolate the effect of data redundancy on kernel performance without inter-node communication overhead.

**PhotoNs-2.0 [46]**: A cosmological N-body simulation for stellar dynamics, based on a non-uniform particle distribution moving over time was developed in [42-45]. It employs an irregular binary MLFMA tree per MPI process to handle load imbalance, with tree rebuilding at each time step. The P2P operator also executes separately for three kind of neighborhoods leading to many small kernel executions. The original implementation [46][48] is MPI-based and lacks GPU acceleration, however their GPU version [46] was not publicly published. We developed a CUDA-enabled version of its near-field (P2P) operator [49], incorporating both baseline (SoA + Indexing[1]) and redundant data layouts. This dynamic, irregular workload without any symmetry in neighboring interactions (which could accelerate particle movement simulations [20]) presents a challenging but realistic scenario for memory-bound GPU kernels.

All experiments were conducted on a workstation equipped with an Intel Core-i7 (7th Gen), 16 GB DDR4 RAM, and an NVIDIA GeForce GTX 1050 (4 GB GDDR5), running both Windows 10 and Ubuntu 24.04.

This single-GPU setup allows us to focus on memory access behavior and locality effects without interference from inter-node communication. Furthermore it enable fair comparison between the static (DBIM) and dynamic (PhotoNs) workloads and validate our analytical model in a controlled, reproducible environment.

We applied data redundancy at the *block-level* for DBIM and at the *thread-level* for PhotoNs, tailored to their respective access patterns. Performance was measured in terms of kernel speedup and end-to-end execution time, with results compared against predictions from our locality-based model.

## 5. Experimental Results

### 5.1. Case Study 1: DBIM-MLFMA [50] – Electromagnetic Inverse Scattering with Uniform Distribution

Mert et al. [50] addressed an inverse electromagnetic scattering problem for tomographic image reconstruction using the Distorted Born Iterative Method (DBIM), an iterative optimization algorithm. In this setup, electromagnetic (EM) cameras are placed around a circular domain to measure the scattered field from an unknown object. By comparing the measured received field with the field computed from an estimated object model, the algorithm iteratively updates the object's properties -specifically its magnetic field absorption coefficient -until convergence to the true object is achieved.

Solving the Helmholtz equation for Green function leads to a nested optimization process, the outer is DBIM and the inner on is BiCGStab algorithm which solves a matrix multiplication of type $XY^{-1}$ and solves it via the (MLFMA). For its P2P operator, all $9t^2$ neighboring patterns (symmetric and repeated distance vectors between particle pairs in E2 neighborhood [4] is calculated once before algorithm execution and is used later by loading into sharing-memory and each thread uses them to calculate all its interactions with its E2 neighbor samples. Each box is divided into t square sub-cells with indicate samples. The value of t is constrained to be a power of 2 (e.g., t=64 ), resulting in dense, uniformly populated boxes. However, on a GPU, this creates a memory contention issue: multiple thread blocks competing to load the shared interaction patterns from the same global memory region into their respective sharing memories.

### 5.1.1. Sensitivity to Density and Tree Structure

The performance of MLFMA is highly sensitive to the density of discretization and the tree depth (level). In this application the tree level and density is fixed and density is controlled by the number of samples per box, t, and the total number of boxes (determined by N and level L ). Three configurations are particularly relevant: Base Tree (Moderate Density), Dense Tree and Less Dense Tree. In base tree $t$ is fixed to 64 (i.e., 8×8 samples per box) and tree level L scaling logarithmically with $N$. This results in a balanced distribution of work between far-field and near-field operators. In less dense tree, the tree depth is increased by one level ($L'=L+1$ ), reducing the number of samples per box to t=16 (i.e., 4×4 ), while keeping N constant. This decreases the size of shared interaction patterns and reduces memory footprint per box. In dense tree (t=256) most of the runtime is bounded to near-field computation totally makes it slower than the base tree because the threads workload increases due to number of neighbours, i.e., each thread processes 16 times larger set of neighbours. The share of Near-Field operator was shown in Figure [6] too.

Figure[3] and Table[1] show the relationship between tree density and runtime components. The left plot shows that near-field runtime increases with t, as larger interaction patterns require more computation. The middle plot shows the inverse behavior for far-field, which benefits from coarser boxes. The right plot reveals that increasing tree level has minimal effect on total MLFMA runtime, but decreasing the level significantly degrades performance.

It suggests that we must focus on improving the less dense tree since the redundancy cant level up the speedup significantly from 0.4 to at east 1. The dominance of the near-field operator increases with box size (t), however, in denser trees (larger t) the fraction of time spent in near-field computation is already high, and the baseline implementation is well-optimized, leaving little chance for improvement via redundancy. Conversely, in less dense trees (smaller t), although the near-field takes less time relatively, the regularity and smaller pattern size make redundancy more effective -because the duplicated data fits better in cache, and dispersion is reduced more efficiently.

| N | Dense Tree L'=L-1, t=256 | | | Base Tree L'=L, t=64 | Less Dense Tree L'=L+1, t=16 | | |
|---|---|---|---|---|---|---|---|
| | P2P Speedup | P2P Share | Overall Speedup | P2P Share | P2P Speedup | P2P Share | Overall Speedup |
| 65536 | 0.3 | 0.75 | 0.41 | 0.47 | 2.81 | 0.13 | 1.05 |
| 262144 | 0.27 | 0.88 | 0.39 | 0.47 | 3.37 | 0.12 | 1.11 |
| 1048576 | 0.28 | 0.82 | 0.41 | 0.49 | 3.35 | 0.13 | 1.02 |
| 4194304 | 0.26 | 0.82 | 0.39 | 0.49 | 3.4 | 0.13 | 1.03 |
| 16777216 | 0.26 | 0.82 | 0.38 | 0.49 | 3.32 | 0.12 | 1.01 |

Table [1] Test speedup of MLFMA implemented in [50] based on density. Ass density decreases, P2P becomes faster while at the same time far-field computation increases, makes it neutral. Base tree level for N=65536 is 5 and for other records increase by one, for N=16777216 it is 9.

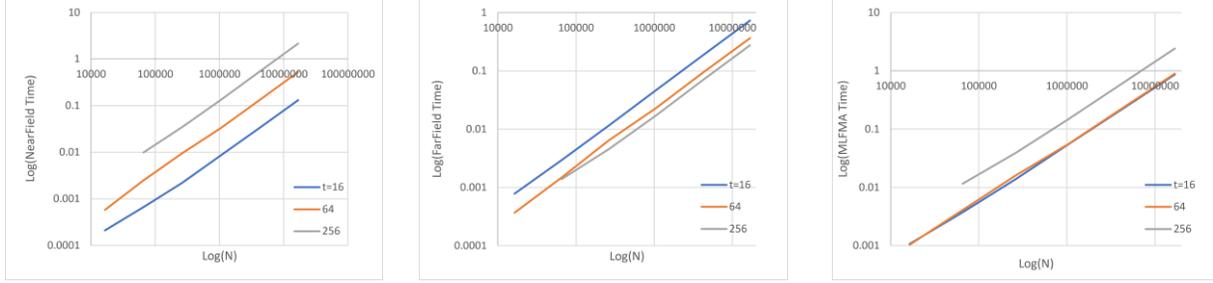

Figure[3] Run time of problems with different N and density t. Density is 64,256,16 for base tree, dense tree and less dense tree respectively.

### 5.1.2. Applying Redundancy

The static and regular nature of this problem limits the potential for dramatic improvements through redundancy. Since the data access pattern is already highly structured and predictable, the baseline implementation already achieves a high degree of memory efficiency.

Applying redundancy in thread level here leads to huge amount of redundant data which makes redundancy not efficient for a thread per particle task distribution. To mitigate this, we apply data redundancy at the block level by replicating the shared interaction pattern in global memory.

We define the *Redundancy Factor (RF)* as the number of copies of the shared pattern structure created in global memory. The goal is to allow each active thread block to access a specific copy of the data for reducing cache contention. Theoretical analysis suggests that RF should match the maximum number of thread blocks that can reside concurrently on a single Streaming Multiprocessor (SM). However, increasing RF increases both the data volume and the overhead of data collection and transfer from host to device. To simplify the modeling process and eliminate RF as a variable in the search space, we fix RF = 2 for all experiments in this study. This choice balances performance gain with manageable overhead and allows us to focus on analyzing the impact of redundancy rather than tuning its degree. Our implementation is published in [52].

### 5.1.2. Modeling the Speedup

The speedup of the modified P2P kernel is modeled using the formula (5),(6) however we ignore data collection and data update components since the baseline implementation [50] reuses precomputed patterns and does not perform data collection at each iteration. There is no further update and data restructuring also required since the new data is replaced in transfer operation. In the redundant version, the data transfer operation considers copy of unknowns and patterns (which is the same for base implementation) plus copy of redundant patterns.

The share of each component was approximated from experimental runs of the baseline code across five different N values and three tree levels. Due to fluctuations and system-level noise, we model the shares as smooth functions of t using a least-square process:

$$Share_{Transfer} \approx 0.5 - 0.087 \ln(t) \tag{12}$$
$$Share_{Compute} \approx 1 - Share_{Transfer} \tag{13}$$

See Figure [4][5][6] for the experimental and approximated trend of share of components and Table [2], Appendix1, for speed-up formulas and their calculations respectively.

| Component | Base method | Redundant method | Speed-up |
|---|---|---|---|
| Transfer | $48N$ | $48N + 288t^2$ | $\dfrac{1}{1 + \dfrac{6t^2}{N}}$ |
| Compute | $32N + 288t^2 + 9Nt^2$ | $32N + RF \times 288t^2 + 9Nt^2$ | $1 + \dfrac{288t^2}{32N + 288t^2 + \dfrac{9N}{t}}$ |

Table [2] Speed-up components for redundancy applied on work [50]

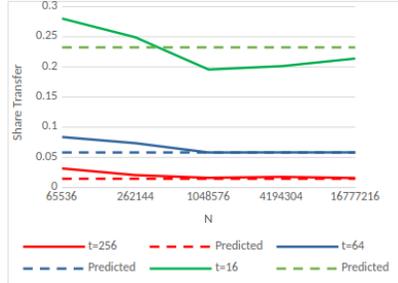 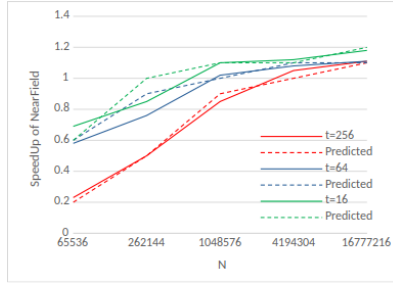 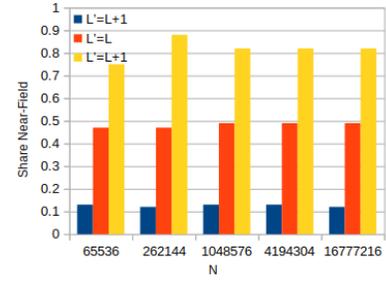

Figure[4] Share of Transfer time for different N and t, with their predicted values.

Figure[5]: Speedup of Near-Field (P2P) for different N,t and their predicted values

Figure[6] Share of Near-Field (P2P) from total execution time for different values of t,N.

Figure [5] compares the predicted speedup trend with measured P2P runtime across different t values. Not only the model capture absolute values with acceptable error rate (average relative absolute error ~9%, correlation ~ 94%), but also it correctly predicts the trend: speedup decreases as t increases, due to imposed overheads.

As shown in Table [3], the GPU kernel achieves a speedup of about 1.2× for smaller problem sizes when redundancy is applied, decreasing to 1.1× for larger problems. However, due to the relatively small contribution of kernel computation to the total P2P phase -and the even smaller share of P2P within the full MLFMA execution -this kernel-level improvement is significantly attenuated at higher levels of the application.

Consequently, the effective end-to-end speedup of the P2P operator ranges from 0.26× to 0.84× for smaller problems and 1.02× to 1.9× for larger ones. When these values are scaled by the ratio of execution time relative to the base-level tree, the results reveal substantial variation but an overall increasing trend with both problem size and tree depth. For large-scale configurations, the total speedup reaches 1.03X, 1.04× for the base tree and the less dense tree respectively.

| | | | | Speed-Up of redundancy to base implementation, same level tree | | | | Speed-Up of redundancy to base implementation, base level tree, for MLFMA |
|---|---|---|---|---|---|---|---|---|
| | N | L | t | MLFMA | Near-Field | Transfer | Kernel | |
| Dense Tree | 65536 | 4 | 256 | 0.26 | 0.23 | 0.01 | 1.2 | 0.11 |
| | 262144 | 5 | 256 | 0.57 | 0.54 | 0.02 | 1.17 | 0.22 |
| | 1048576 | 6 | 256 | 0.88 | 0.87 | 0.05 | 1.14 | 0.36 |
| | 4194304 | 7 | 256 | 1.03 | 1.04 | 0.21 | 1.12 | 0.4 |

|  | N | L | t |  |  |  |  |  |
|---|---|---|---|---|---|---|---|---|
|  | 16777216 | 8 | 256 | 1.09 | 1.11 | 0.54 | 1.12 | 0.41 |
| Base Tree | 65536 | 5 | 64 | 0.65 | 0.55 | 0.08 | 1.24 | 1.09 |
|  | 262144 | 6 | 64 | 0.92 | 0.88 | 0.23 | 1.14 | 0.68 |
|  | 1048576 | 7 | 64 | 1.01 | 1.03 | 0.49 | 1.1 | 0.79 |
|  | 4194304 | 8 | 64 | 1.05 | 1.09 | 0.93 | 1.11 | 1 |
|  | 16777216 | 9 | 64 | 1.06 | 1.11 | 1.29 | 1.1 | 1.04 |
| Less Dense Tree | 65536 | 6 | 16 | 0.84 | 0.61 | 0.28 | 1.16 | 0.88 |
|  | 262144 | 7 | 16 | 0.98 | 0.96 | 0.7 | 1.08 | 1.09 |
|  | 1048576 | 8 | 16 | 1.01 | 1.06 | 0.93 | 1.09 | 1.03 |
|  | 4194304 | 9 | 16 | 1.02 | 1.12 | 1.28 | 1.09 | 1.05 |
|  | 16777216 | 10 | 16 | 1.02 | 1.16 | 1.57 | 1.08 | 1.03 |

Table[3] Speed-up components for redundancy applied on work [50]

### 5.1.3. Overall Speed-up of the Application

To assess the real-world impact of our optimization, we integrated the redundant P2P operator into a full DBIM simulation loop. The inner BiCGStab solver requires up to 300 iterations, each involving three MLFMA calls for matrix-vector products. In the original work [50], this process is executed across 16 computing nodes, but at only 1 node it leads to up to 9900 MLFMA invocations.

In our setup, we run a single-node version with 3 DBIM iterations (fixed for comparison), and compute the average execution time per MLFMA call over the entire DBIM loop. Input vectors are generated randomly to eliminate bias from convergence behavior. The results are summarized in Table [4] showing that for the whole application where less data transfer occurs, the total speed-up for large problems reaches to 5% for base level tree and 10% for less dense tree, however it reduces to 4% when compared to base level tree. This shows that changes in density didn't improve the overall speedup more than the base level tree. We define 5%X as the net speedup of the P2P operator under data redundancy. Importantly, this table does not represent the prediction of our analytical model - it shows the actual end-to-end speedup achieved in the full DBIM application.

| N | L | t | Average Running Time MLFMA [50] | Average RunningTime MLFMA [50] +P2P Redundancy | Speedup Redundancy | Speedup Redundant to Base Level Tree |
|---|---|---|---|---|---|---|
| L'=L | | | | | | |
| 262144 | 6 | 64 | 14.88 | 14.4 | 1.03 | |
| 1048576 | 7 | 64 | 56.94 | 55.49 | 1.03 | |
| 4194304 | 8 | 64 | 228.02 | 216.77 | 1.05 | |
| L'=L+1 | | | | | | |
| 262144 | 7 | 16 | 15.6 | 15.14 | 1.03 | 0.98 |
| 1048576 | 8 | 16 | 63.67 | 56.12 | 1.1 | 1.01 |
| 4194304 | 9 | 16 | 242.33 | 219.5 | 1.1 | 1.04 |

Table [4]: Average speedup of MLFMA with P2P redundancy in the DBIM algorithm.

### 5.1.4. Conclusion on EM Case Study

Despite the highly optimized, static, and regular nature of the DBIM problem, our block-level redundancy technique achieves up to 10% kernel speedup in less dense configurations.

However, due to the dominance of precomputed patterns, efficient memory access in the baseline, and increased data transfer overhead, overall end-to-end speedup remains limited (~4%).

The work of [50] was originally designed for large GPU clusters, where inter-node synchronization and communication dominate. In such environments, the independence introduced by redundant data could eliminate synchronization bottlenecks, potentially yielding greater benefits than observed here on a single node. Thus, while redundancy does not dramatically improve performance in this specific single-GPU, it demonstrates a viable path for future optimization in distributed settings, and validates our locality-based modeling approach in predicting performance trends.

## 5.2. Case Study 2: PhotoNs-2.0 [NS 3] – Non-Uniform Particle Distribution with Irregular Binary Tree

The PhotoNs-2.0 code [48] is a cosmological N-body simulation framework designed for large-scale gravitational dynamics. It combines Particle-Mesh (PM) and hierarchical tree methods (PM-Tree) to efficiently compute long-range and short-range forces, respectively. While the original implementation is CPU-based and relies on MPI for parallelization, no publicly available GPU-accelerated version had publicly. Therefore, we developed a CUDA-based GPU implementation of the near-field (P2P) operator to evaluate the impact of data redundancy on performance in dynamic, irregular astrophysical simulations.

Interactions are categorized as local, remote, or periodic (to handle toroidal boundary conditions in 3D space). Periodic interactions alone require 27 (E2 neighbors) separate kernel launches per iteration, in addition to one for local and at least one for remote interactions. For modeling the speedup, we only consider local interactions which is wasy to analyse and predictable. The small invocations surpass a static overhead to execution time which is not counted in our model.

The binary tree structure is highly irregular, with varying depths and unbalanced leaf distributions across sub-domains. This leads to load imbalance and non-uniform memory access patterns. Further more no symmetric or shared interaction patterns exists and pre-computations are not applied here.

### 5.2.1. Applying Redundancy

We implemented two variants of the P2P operator in [49]:

Indexing Implementation: Based on the approach described in [1], this SoA version uses a Structure-of-Arrays (SoA) layout. Each thread processes one target particle and computes its interactions with all source particles in neighboring boxes. Particle indices are first loaded, followed by data access from global memory, interaction computation, and result accumulation.

Redundant Implementation: Applies thread-level data redundancy by duplicating particle data for each interaction pair. The data is restructured into an Array-of-Structures (AoS) format, where each entry contains both source and target attributes. This enhances spatial locality, allowing threads to access all required data from a compact, contiguous memory block, thereby reducing cache misses.

Both implementations were compared against the original MPI-based baseline [46], executed with 4 processes on our test system, they only rely on GPU for acceleration. For this application we ignore evaluation of different tree levels unlike section 4.1 since for [46] there are many MLFMA trees with different levels for each iteration.

### 5.2.2. Experimental Results

We evaluated performance over 500 iterations for clustering thresholds t=2,4,8,16,64 . Figure [Figure-P1] compares total simulation time, breakdown of baseline execution, and P2P computation time. We generated five key plots to analyze the performance in Figure [7].

Figure [7].a compares the total simulation time across the three implementations. Figure [7].b compares the P2P operator runtime (data collection, transfer, compute, update) for all three versions. For higher particle densities, the total simulation time increases nearly exponentially, with the P2P operator being the dominant contributor. In the GPU implementation, the overall runtime decreases sub-linearly up to t=16 , but then gradually increases for larger t. Despite these optimizations, the Redundant implementation does not outperform the Indexing implementation in terms of overall execution time.

Figure [7].d reports the speedup of different P2P interaction types (local, remote, periodic) in the Redundant implementation, showing how remote and periodic computations slow down by increase in density. Figure [7].e breaks down the speedup of individual P2P components (collection, transfer, compute, update) and shows that the kernel computations reaches to 3X faster in Redundant implementation for t=8 (default value) and always faster than Indexing implementation also the update process which is faster because of its data volume. The Indexing implementation consume larger memory volume due to its padding scheme and the non-uniform particle distribution.

That the P2P speedup for Redundant implementation raises from 0.55X for t=2 to 0.9X for t=32 and drops to 0.81X for t=64 as depicted in Figure [7].f .The overall applications speedup grows gradually from 0.79X for t=2 to 0.98X for t=32 and drops to 0.94X for t=64 which is better than the P2P speedup rooted in reduced share of P2P in overall runtime.

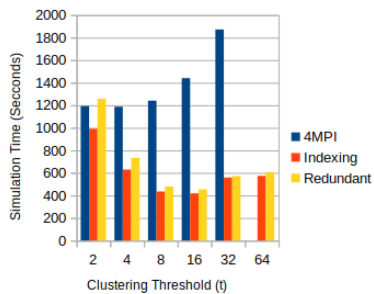

a

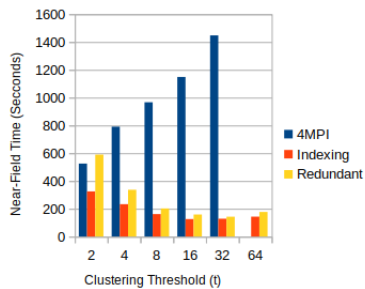

b

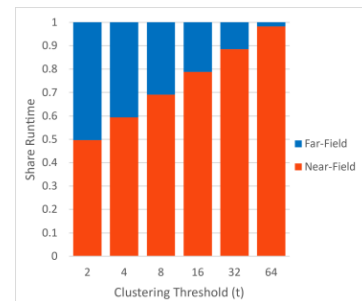

c

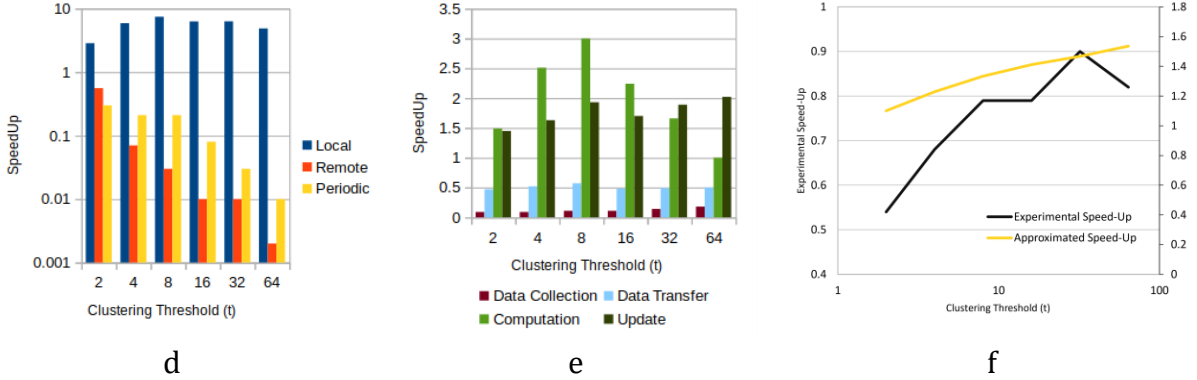

F*igure* [7]: Comparing baseline implementation of Photo-Ns[45] with GPU implementations. The application was ran with data-set [48] on our system with 4 MPI processes for five different density thresholds (2...64) but for MPI implementation t=64 could not be executed. **a-** Comparison of total simulation time. **b-** Comparison of Near-Field or P2P execution time. **c-** Break-down plot for execution time of 4 MPI baseline implementation. **d-** speedup for different types of interactions. **e-** speedup of each runtime component. **f-**overall speedup of P2P and its prediction using model.

### Local Computation Shows High Kernel Speedup

according to Figure [7].d the most significant performance gain is observed in local P2P computations, where data redundancy improves spatial locality and reduces cache misses. For t=32, the speedup of the local interaction kernel averages 6.3×, increasing from 4.4× in the first iteration to 7.4× to the last iteration. This improvement is attributed to the compact, contiguous memory access pattern enabled by the redundant AoS layout.

### Speedup Reduction Across Execution Layers

As shown in Figure [7].d, the high speedup in local kernel computation is significantly reduced when higher-level components are considered. For t=8, the local kernel achieves a 7.4× speedup, but when combined with remote and periodic interactions, the total P2P compute speedup drops to 3X and when combined with other components it drops to ~0.79× as shwn in Figure [7].f. Finally, when accounting for the fraction of P2P time in the total simulation (Figure [7].c), the overall simulation speedup reaches to 0.9×. This progressive reduction demonstrates how local gains are diluted by system-level overheads.

### Optimal Performance at t=32

Among all tested configurations, t=32 emerges as the balance point between kernel speedup and overhead. At this threshold where the local kernel achieves a high speedup (6.3×), and at the same time the fraction of P2P time in total execution is 0.85 as shown in Figure [7].c. The overheads from collection and transfer, while present, do not fully negate the compute gains.

As a result, the overall speedup peaks at this point, with the redundant implementation achieving a total simulation speedup close to 0.98× (i.e., a 2% slowdown), which is the best observed performance across all t values. For t=64, although data volume increases further, the number of interactions decreases, and the P2P share of total time diminishes, leading to even lower relative gains.

### 5.2.3. Modeling the SpeedUp

Several inherent characteristics of the PhotoNs-2.0 application introduce significant challenges for both optimization and analytical modeling:

1. **Dynamic Data Restructuring**: Particle positions evolve at each time step, requiring repeated data collection and restructuring before every P2P call. Overhead of data-restructuring particularly in the redundant version where data duplication must be redone iteratively, offset the redundant kernel speedup.
2. **Irregular and Imbalanced FMM Trees**: This limit the effectiveness of padding and alignment optimizations. Further more no symmetric or shared interaction patterns exists and pre-computations are not applied here. One side effect of varying trees is that the number of key elements for runtime prediction like number of E2 interactions and number of final level boxes (leafs) is unknown. In Figure[8] the dynamic of these variables for a single simulation or multiple values of t is depicted. We approximated their trend as a smooth function of t for modeling in Appendix 2.
3. **Decomposition of Interactions**: Figure [9] depicts the gap between speedup of three types of interactions, showing the negative effect of redundancy for several small sized kernels execution. This suggests an update for (formula4) where the share of three types of interactions are accounted as (formula 14),(formula 15) and need to calculate the average $Share_{localC}$ from experimental results.

$$SpeedUp_{ComponentC} = \frac{Share_{LocalC}}{SpeedUP_{LocalC}} + (1 - Share_{LocalC}) \text{ (formula14)}$$

$$Share_{LocalC} = \frac{Time_{Local}}{Time_{Local} + Time_{Remote} + Time_{Periodic}} \text{ (formula15)}$$

4. **Numerous Small Kernel Launches**: Due to GPU memory limitations and the original queue size constraint (16,384 tasks), we process interactions in batches. In the indexing version, each interaction type is processed in a single kernel call. In the redundant version, due to higher memory footprint, we invoke the kernel in batches of ~20,000 tasks, leading to frequent small launches.
5. **Overlapping Computations**: The original code overlaps M2L and P2P computations to hide latency. However, this complicates modeling due to timing dependencies. To simplify analysis, we disabled overlapping and executed all P2P computations after M2L completion.
6. **Dispersion effect**: Increased data dispersion did not lead to a reduction in cache misses due to the larger size of the data each thread fetches. As a result, we adopt Formula (11) in place of Formula (10) for modeling kernel speedup, as it better reflects the observed behavior. Interestingly, dispersion reduced overall data volume, primarily because it avoids the use of padding schemes commonly employed in Indexing scheme.

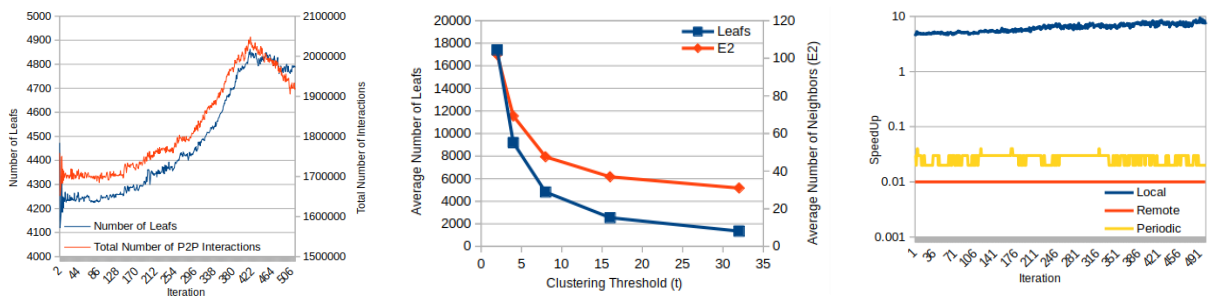

Figure [8]: Irregular tree parameters for Photo-Ns [**NS3**]. Left: Varying number of tree leafs (last level boxes) and P2P interactions over simulation iterations for clustering threshold (t) = 2. Right: Average number of leafs and E2 neighbors based on clustering threshold.

Figure[9]: Speed-Up of different *interactions* for a simulation with t=32. The speed-Up of local computation is The modeling of runtime components are summarized in Table [5] and Appendix2, which compares complexity and memory usage for local interactions only. The share of each component in the baseline (indexing) implementation is approximated from experimental data across multiple problem sizes using a least square technique and function of clustering thresholds t. As shown in Table [6] we model these shares as smooth analytical functions rather than exact piecewise expressions due to system noise and non-uniform behavior across t, avoiding unnecessary complexity while preserving trend fidelity. The share of components for Indexing implementation is depicted in Figure [10] separately.

|  | Indexing Scheme | Redundant Scheme | X (SpeedUp) |
|---|---|---|---|
| Data Collection Complexity | $3 \cdot leafs \cdot t + 2 \cdot interactions$ | $6t \cdot interactions$ | $21E2 + 3t1$ |
| Data Collection Memory | $leafs(152t + 24394)$ | $leafs(24E2(1 + 4t) + 104t + 8376)$ | $24E2(1 + 4t) + 104t + 8376152t + 24394$ |
| Transfer Memory | $16 \cdot leafs(3t + 1 + Max\_E2 + 3t \cdot Max\_E2)$ | $8 \cdot interactions(9t + 1.5)$ | $E2(9t + 1.5)2(3t + 1001 + 3000t)$ |
| Kernel Complexity | $4 + 6t + 3t2$ | $2 + 6t + 3t2$ | $1$ |
| Kernel Dispersion | $t2 + 2t + 2$ | $t + 643t(t + 1) + 1$ | $643t(t + 1) + t + 1t2 + 2t + 2$ |
| Kernel Launch Count | $1 + num\_remote\_launch + num\_periodic\_launch$ | Same | $1$ |
| Update Complexity | $interactions(2 + 15t)$ | Same | $1$ |
| Update Memory | $8 \cdot interactions + leafs(376 + 104E2 + 48000t)$ | $8 \cdot interactions + leafs(376 + 24E2)$ | $< 3$ |

Table [5]: Complexity and memory usage comparison between indexing and redundant implementations for local P2P computations. Definitions: E2 = number of E2 interactions per leaf; t = number of samples per dimension; num_remote_launch, num_periodic_launch : number of kernel launches for remote and periodic interactions.

| Component | Share runtime |
|---|---|
| Data Collection | $0.05 - 0.005\ln(t)$ |
| Data Transfer | $0.5 - 0.087\ln(t)$ |
| Kernel | $0.18\ln(t)$ |
| Update | $0.5 - 0.11\ln(t)$ |
| Near-Field (Total) | $0.4 + 0.14\ln(t)$ |

Table [6]: Approximation of time share for each component in the indexing implementation based on experimental data.

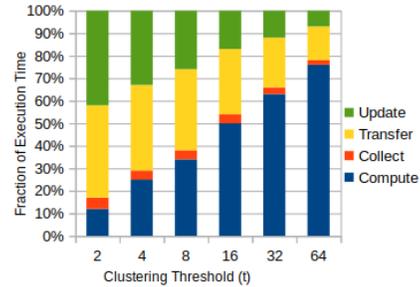

Figure [10]: Fraction of each step of computation in Indexing implementation for different leaf size (t)

Figure [7].f compares the predicted speedup trend with experimental results. Despite capturing the general decline in speedup with increasing t, the model shows average absolute relative error ~80% but correlation ~92%. The discrepancy arises from unmodeled system overheads -especially kernel launch latency and dynamic restructuring costs. Nevertheless, the model correctly predicts performance trends, validating its utility for early-stage design guidance.

### 5.2.4. Conclusion on Stellar Dynamics Case Study

PhotoNs-2.0 represents a realistic, dynamic, and challenging workload for GPU acceleration. While thread-level redundancy improves kernel performance by up to 7X, system-level overheads -particularly numerous small kernel launches and dynamic data restructuring -limit end-to-end gains and reduce the speedup to about 0.9X.

The model, though inaccurate in absolute prediction and predicts on average 1.8 times larger values for speedup, successfully captures the trend, demonstrating that data redundancy is most beneficial when locality is poor and kernel execution dominates. In highly fragmented, communication-heavy applications like PhotoNs, optimizing the execution pipeline (e.g., batching, overlapping) is as critical as kernel-level optimization. The alignment of predicted and observed trends confirms that data locality and memory access patterns are critical factors in achieving speedup through data restructuring on GPUs.

Although we lacked access to the GPU implementation in [48], our design appears more memory-efficient. Figure [11] compares memory access patterns. In [48], tasks are grouped by source leaf to avoid write races, with one thread handling multiple interactions. However, target leaf data is scattered in memory, leading to non-coalesced accesses -even within groups -since spatial adjacency is not guaranteed. In contrast, our Indexing implementation assigns one thread per interaction, storing particle data in separate arrays and deferring updates to CPU, but coalescing remains suboptimal (e.g., repeated accesses to the same leaf). In our Redundant implementation, duplicated data enables threads to access contiguous blocks, significantly improving spatial locality and achieving coalesced memory access.

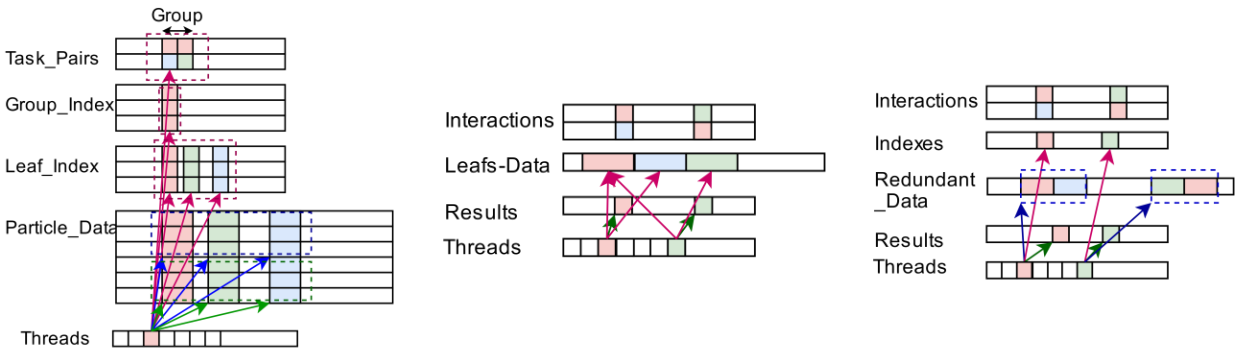

Figure [11] – comparing memory access patterns in three implementations: left [46] Indexing and Redundant implementations (middle and right) [49]

## 6. Conclusion And Future Works

This work presents a foundational framework for understanding how data restructuring, specifically redundancy, impacts GPU performance in MLFMA. By introducing an analytical model based on spatial locality, we aim to enable an early evaluation of optimization strategies, however it requires some experimental measurements of baseline algorithm and needs to consider memory block size. The model, while not perfectly predictive, reliably identifies performance trends and helps determine when redundancy is beneficial.

Applied to electromagnetic, astrophysical, and potential-function problems, our approach achieved up to 7× kernel speedup and up to 4% end-to-end improvement, despite data transfer overheads.

The results show that speedup increases for denser problems, provided that the far-field computations do not become a bottleneck due to their growing share of execution time (as observed in the DBIM case), and as long as the data volume remains within limits where cache benefits are still effective (unlike in the stellar dynamics case, where high data volume diminishes caching advantages). For the first case, the results are consistent with the model's predictions, demonstrating its ability to accurately capture performance trends under controlled, static conditions. However, for the second case, the model fails to fully predict the return in speedup due to the complex and dynamic runtime behavior of the code, including frequent kernel launches, irregular data access patterns, and system-level overheads that are difficult to abstract analytically.

Future work will focus on extending the proposed model to multi-GPU systems and investigating automated methods for locality estimation. If the framework is validated across standard GPU benchmarks or restricted to stencil-based applications, it could serve as a foundation for automating data restructuring decisions to enhance parallel efficiency in large-scale simulations.